\documentclass[a4paper,11pt]{article}
\usepackage{pos}

\title{Machine learning in LHCb Simulation: From fast to flash}
\manuallySeparateAuthors
\author*[a]{Michał Mazurek}
\author{on behalf of the LHCb Simulation Project}

\affiliation[a]{National Centre for Nuclear Research,\\
  Andrzeja Sołtana 7, Otwock, Poland}

\emailAdd{michal.mazurek@cern.ch}

\usepackage{hyperref}
\usepackage{caption}
\usepackage{subcaption}
\usepackage{float}

\usepackage{ifthen} 
\newboolean{uprightparticles}
\setboolean{uprightparticles}{false} 
\usepackage{xspace} 
\usepackage{upgreek}

\usepackage{xspace} 
\usepackage{upgreek}

\def\lhcb   {\mbox{LHCb}\xspace}

\def\rich   {RICH\xspace}

\def\MagUp {\mbox{\em Mag\kern -0.05em Up}\xspace}

\ifthenelse{\boolean{uprightparticles}}%
{

 \def\PDelta      {\ensuremath{\Delta}\xspace}                 
 \def\PXi         {\ensuremath{\Xi}\xspace}                 
 \def\PLambda     {\ensuremath{\Lambda}\xspace}                 
 \def\PSigma      {\ensuremath{\Sigma}\xspace}                 
 \def\POmega      {\ensuremath{\Omega}\xspace}                 
 \def\PUpsilon    {\ensuremath{\Upsilon}\xspace}
 \let\oldPi\Pi
 \def\PPi         {\ensuremath{\oldPi}\xspace}

 \def\PB      {\ensuremath{\mathrm{B}}\xspace}                 
 \def\PD      {\ensuremath{\mathrm{D}}\xspace}                 

 \def\PK      {\ensuremath{\mathrm{K}}\xspace}                 
 \def\Ps      {\ensuremath{\mathrm{s}}\xspace}

 \def\thebaroffset{0.0em}
}
{

 \mathchardef\PDelta="7101
 \mathchardef\PXi="7104
 \mathchardef\PLambda="7103
 \mathchardef\PSigma="7106
 \mathchardef\POmega="710A
 \mathchardef\PUpsilon="7107
 \mathchardef\PPi="7105
 \def\PB      {\ensuremath{B}\xspace}                 
 \def\PD      {\ensuremath{D}\xspace}                 

 \def\PK      {\ensuremath{K}\xspace}                 
 \def\Ps      {\ensuremath{s}\xspace}

 \def\thebaroffset{0.18em}
}
\newcommand{\offsetoverline}[2][\thebaroffset]{\kern #1\overline{\kern -#1 #2}}%

\makeatletter
\ifcase \@ptsize \relax
  \newcommand{\miniscule}{\@setfontsize\miniscule{4}{5}}
\or
  \newcommand{\miniscule}{\@setfontsize\miniscule{5}{6}}
\or
  \newcommand{\miniscule}{\@setfontsize\miniscule{5}{6}}
\fi
\makeatother

\DeclareRobustCommand{\optbar}[1]{\shortstack{{\miniscule (\rule[.5ex]{1.25em}{.18mm})}
  \\ [-.7ex] $#1$}}

\def\squark    {{\ensuremath{\Ps}}\xspace}

\def\KorKbar {\kern \thebaroffset\optbar{\kern -\thebaroffset \PK}{}\xspace}

\def\D       {{\ensuremath{\PD}}\xspace}

\def\DorDbar {\kern \thebaroffset\optbar{\kern -\thebaroffset \PD}\xspace}

\def\Dp      {{\ensuremath{\D^+}}\xspace}
\def\Dm      {{\ensuremath{\D^-}}\xspace}

\def\DpDm    {\ensuremath{\Dp {\kern -0.16em \Dm}}\xspace}

\def\B       {{\ensuremath{\PB}}\xspace}

\def\BorBbar {\kern \thebaroffset\optbar{\kern -\thebaroffset \PB}\xspace}

\def\Bd      {{\ensuremath{\B^0}}\xspace}

\def\BdorBdbar {\kern \thebaroffset\optbar{\kern -\thebaroffset \Bd}\xspace}

\def\Bs      {{\ensuremath{\B^0_\squark}}\xspace}

\def\BsorBsbar {\kern \thebaroffset\optbar{\kern -\thebaroffset \Bs}\xspace}

\def\Y#1S{\ensuremath{\PUpsilon{(#1S)}}\xspace}

\def\LorLbar     {\kern \thebaroffset\optbar{\kern -\thebaroffset \PLambda}\xspace}

\def\to                 {\ensuremath{\rightarrow}\xspace}

\def\AT#1     {\ensuremath{A_{\mathrm{T}}^{#1}}\xspace}           

\def\C#1      {\ensuremath{\mathcal{C}_{#1}}\xspace}                       
\def\Cp#1     {\ensuremath{\mathcal{C}_{#1}^{'}}\xspace}                    
\def\Ceff#1   {\ensuremath{\mathcal{C}_{#1}^{\mathrm{(eff)}}}\xspace}        
\def\Cpeff#1  {\ensuremath{\mathcal{C}_{#1}^{'\mathrm{(eff)}}}\xspace}       
\def\Ope#1    {\ensuremath{\mathcal{O}_{#1}}\xspace}                       
\def\Opep#1   {\ensuremath{\mathcal{O}_{#1}^{'}}\xspace}                    

\newcommand{\aunit}[1]{\ensuremath{\text{\,#1}}}       

\newcommand{\tev}{\aunit{Te\kern -0.1em V}\xspace}
\newcommand{\gev}{\aunit{Ge\kern -0.1em V}\xspace}
\newcommand{\mev}{\aunit{Me\kern -0.1em V}\xspace}
\newcommand{\kev}{\aunit{ke\kern -0.1em V}\xspace}
\newcommand{\ev}{\aunit{e\kern -0.1em V}\xspace}
 
\newcommand{\mevc}{\ensuremath{\aunit{Me\kern -0.1em V\!/}c}\xspace}
\newcommand{\gevc}{\ensuremath{\aunit{Ge\kern -0.1em V\!/}c}\xspace}
\newcommand{\mevcc}{\ensuremath{\aunit{Me\kern -0.1em V\!/}c^2}\xspace}
\newcommand{\gevcc}{\ensuremath{\aunit{Ge\kern -0.1em V\!/}c^2}\xspace}

\def\gsim{{~\raise.15em\hbox{$>$}\kern-.85em
          \lower.35em\hbox{$\sim$}~}\xspace}
\def\lsim{{~\raise.15em\hbox{$<$}\kern-.85em
          \lower.35em\hbox{$\sim$}~}\xspace}

\def\gaudi      {\mbox{\textsc{Gaudi}}\xspace}
\def\gauss      {\mbox{\textsc{Gauss}}\xspace}
\def\geant      {\mbox{\textsc{Geant4}}\xspace}
\def\lamarr     {\mbox{\textsc{Lamarr}}\xspace}

\def\pythia     {\mbox{\textsc{Pythia}}\xspace}

\def\tell1  {TELL1\xspace}
\def\ukl1   {UKL1\xspace}

\newcommand{\lhcborcid}[1]{\href{https://orcid.org/#1}{\hspace*{0.1em}\raisebox{-0.45ex}{\includegraphics[width=1em]{LHCb/Figures/orcidIcon.pdf}}}}

\abstract{
Monte Carlo simulations are essential for physics analyses in high-energy physics, but their computational demands are continuously increasing.
In \lhcb, 90\% of computing resources are used for simulations, with the calorimeter simulation being the most computationally intensive part.
Fast simulations and flash simulations, leveraging machine learning techniques, offer promising solutions to this challenge with different levels of detail and speed.
The \textsc{CaloML} framework accelerates electromagnetic shower propagation of photons and electrons in the \lhcb calorimeter by up to two orders of magnitude, achieving a systematic error on reconstructed energies as low as 0.01\%.
\textsc{Lamarr} is an in-house flash simulation framework that reduces CPU time of the whole simulation phase by two orders of magnitude compared to traditional \geant-based methods.
In this paper, these two approaches are presented, highlighting their methodologies, performance, and validation results, as well as future development plans.
}

\FullConference{The Thirteenth Annual Large Hadron Collider Physics (LHCP2025)\\
5-9 May 2025\\
Taipei, Taiwan\\}

\renewcommand{\logo}{\relax}

\begin{document}
\maketitle

\section{Introduction}

The \lhcb detector~\cite{LHCb-DP-2008-001, LHCb-DP-2014-002} is a single-arm forward
spectrometer covering a pseudorapidity range of 2 < $\eta$ < 5, originally designed to study the properties of particles containing beauty (b) or charm (c) quarks.
Its physics program has been extended since the first data taking period to include a wide range of measurements beyond  the field of heavy-flavor physics.
The detector features a high-precision tracking system that measures the momentum of charged particles, an advanced particle identification system, which combines the responses of two Ring-Imaging Cherenkov (\rich) detectors, the calorimeter system, and the MUON system to effectively distinguish between photons, electrons, long-lived hadrons, and muons.

Monte Carlo simulations are crucial for physics analyses in high-energy physics.
\gauss~\cite{LHCb-PROC-2011-006} is the simulation framework used by the \lhcb experiment to handle particle generation and their interactions with the detector.
The demand for simulated samples already limits the precision of certain measurements and is expected to grow.
Since Run~2, simulations have consistently used more than 90\% of the computing resources allocated to the experiment.
To address this, \gauss has been redesigned~\cite{LHCb-TDR-017} to meet statistical requirements for Run~3 and beyond~\cite{LHCb-TDR-018}.
A core framework, \textsc{Gaussino}~\cite{Gaussino:main, Gaussino:main2, Gaussino:main3}, was extracted as a standalone library, enabling multi-threading in the Gaudi framework~\cite{Gaudi:main,Gaudi:main2,Gaudi:main:upgrade} and \geant~\cite{Agostinelli:2002hh} for efficient detector simulation.

Despite the improvements in the simulation framework, the computational cost of simulating the transport and particle interactions with the detector remains a significant challenge.
In particular, the calorimeter simulation is the most computationally intensive part of the simulation process, accounting for  up to 60\% of the total CPU time.
Various approaches have been proposed to lower the computational demands of the simulation phase, including resampling techniques~\cite{LHCb-DP-2018-004} and parameterizations of energy deposits~\cite{FastSimulations:PointLib,FastSimulations:PointLib2,FastSimulations:Fedor,FastSimulations:Rogachev,FastSimulations:ACAT2024,FastSimulations:CHEP2024}.
These methods, collectively referred to as \textit{fast simulations}, provide cost-effective alternatives for replicating the low-level response of the \lhcb detector.
\textsc{CaloML} is the first production-ready, fast simulation option based on generative models for the electromagnetic calorimeter and is described in more detail in Section~\ref{sec:fast-sim}.
An even more drastic approach is represented by \textit{flash simulation} options, which aim to directly parameterize the high-level response of the \lhcb detector.
\textsc{Lamarr}~\cite{FastSimulations:Lamarr,FastSimulations:Lamarr2,FastSimulations:Lamarr3} is an in-house flash simulation based on generative models and is described in more detail in Section~\ref{sec:flash-sim}.

\section{Fast simulations with machine learning}
\label{sec:fast-sim}

Fast simulations aim to replace the most computationally intensive parts of the simulation with fast parameterizations, while still providing a realistic representation of the detector response.
Improvements~\cite{Gaussino:FastSimInterface,Gaussino:FastSimInterface2,Gaussino:FastSimInterface3} in the \textsc{Gaussino} framework have enabled the integration of ML-based fast simulation options as part of the standard simulation workflow. 

\textsc{CaloML}~\cite{FastSimulations:ACAT2024,FastSimulations:CHEP2024,Gaussino:thesis} is the first production-ready fast simulation option using generative models to replace the detailed simulation of the electromagnetic showers inside the \lhcb calorimeter.
It is based on \textsc{CaloChallenge}~\cite{CaloChallenge}, which is the first community-wide challenge for the development of fast and accurate calorimeter simulations.
In the challenge setup, the energy deposits coming from the electromagnetic showers are recorded in virtual concentric cylinders.
These cylinders, dynamically created along the particle's trajectory, are segmented in axial, radial, and azimuthal coordinates.
The \textsc{CaloML} employs similar cylinders in the \textsc{Gaussino} framework, but tailored to the geometry of the \lhcb calorimeter.
Particles are stopped just in front of the calorimeter and their information is stored in a simplified format for further processing.
Once the main simulation algorithm is complete, then the particle information is passed to another \textsc{Gaudi} algorithm for ML-based inference.

Variational Autoencoders (VAEs) were the first models used in the \textsc{CaloChallenge} to generate calorimeter energy deposits and were chosen for preliminary studies in \textsc{CaloML}. To improve the quality of generated energy deposits, a modified VAE model (\textsc{VAEWithProfiles}) was introduced. 
Instead of generating energy deposits directly, the model predicts spatial and energy profiles of the cylinders, which resulted in improvements in both accuracy and training speed.
Additional adjustments account for the calorimeter's non-uniformity, such as passive materials and geometric complexities.

Simulation with the \textsc{CaloML} option achieves up to 2 orders of magnitude faster simulation for electrons and photons compared to the traditional approach. It captures around 40\% of all energy deposits in realistic events, with some limitations in specific regions and particle types. 
With additional tuning of the model, the systematic error of the model on reconstructed energies was reduced to 0.01\% to achieve better agreement between the fast and full simulation scenarios.

\begin{figure}[htbp]
  \centering
  \begin{subfigure}[t]{0.49\textwidth}
    \includegraphics[width=\textwidth]{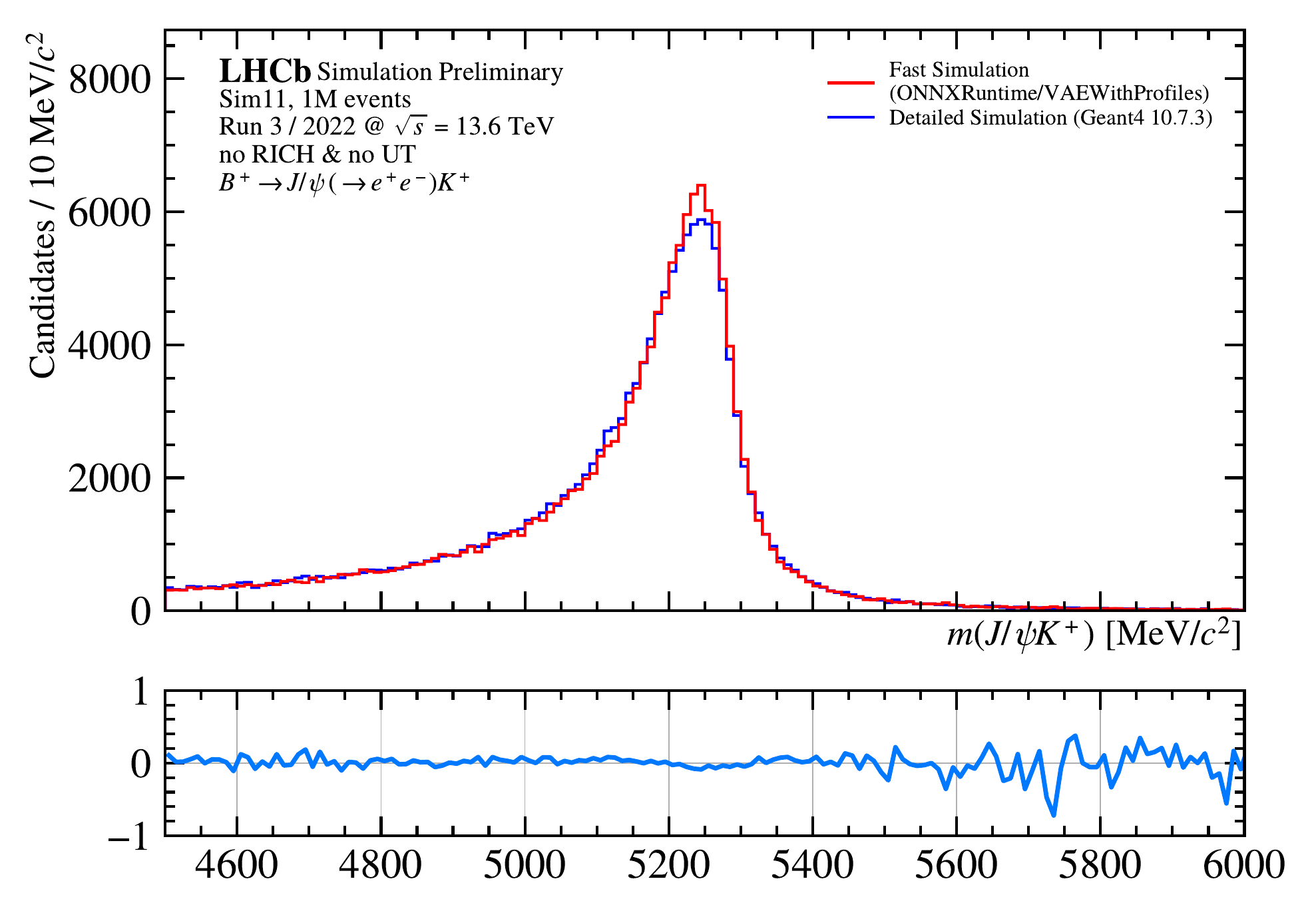}
    \caption{
      $B^+$ invariant mass distribution from the simulation sample of the $B^+ \to J/\psi\;(\to e^+e^-) K^+$ decay.
    }
    \label{fig:VALIDATION_12153001_BU_MASS_DISTRIBUTION}
  \end{subfigure}
  \hfill
  \begin{subfigure}[t]{0.49\textwidth}
    \includegraphics[width=\textwidth]{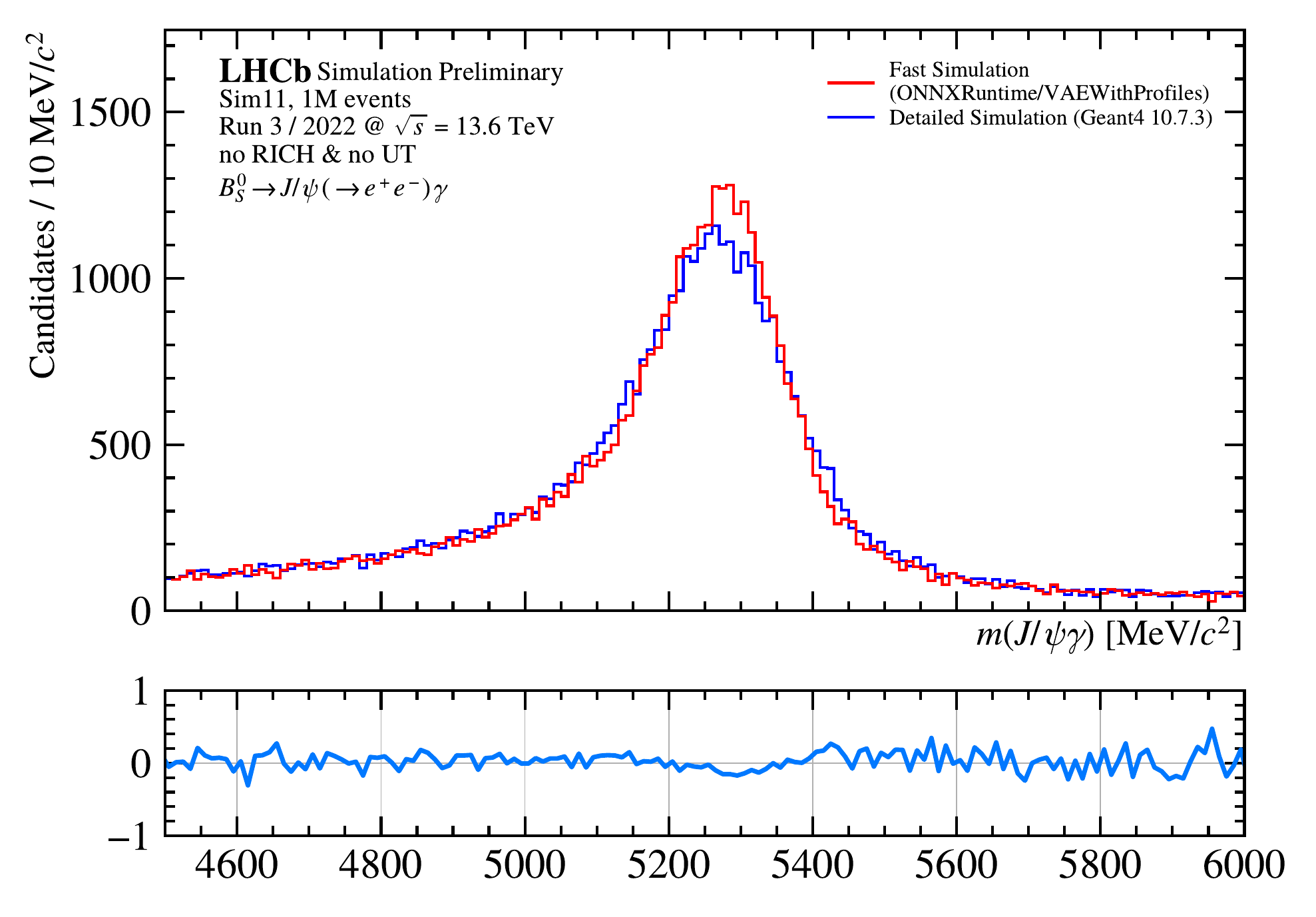}
    \caption{
      $B_s^0$ invariant mass distribution from the simulation sample of the $B_s^0 \to J/\psi (\to e^+e^-) \gamma$ decay.
    }
    \label{fig:VALIDATION_13152201_B0S_MASS_DISTRIBUTION}
  \end{subfigure}
  \caption{Preliminary physics validation of the \textsc{CaloML} fast simulation using two benchmark decay channels.}
  \label{fig:VALIDATION_MASS_DISTRIBUTIONS}
\end{figure}

Physics validation of the \textsc{CaloML} fast simulation demonstrates its ability to reproduce detailed simulation results with high fidelity.
The $B^+$ meson invariant mass distribution (Figure~\ref{fig:VALIDATION_12153001_BU_MASS_DISTRIBUTION}) and $B_s^0$ invariant mass distribution (Figure~\ref{fig:VALIDATION_13152201_B0S_MASS_DISTRIBUTION}) show excellent agreement between the fast and detailed simulation scenarios.
Both distributions are nearly indistinguishable, with only a slight overshoot observed around the mass peak in the fast simulation scenario around the $B_s^0$ meson mass peak.
These results highlight the importance of detailed physics validation to ensure that fast simulation samples closely match those from \geant, as reconstruction algorithms and trigger selections are highly sensitive to simulation quality.

\section{Flash simulations with machine learning}
\label{sec:flash-sim}

\begin{figure}[htbp]
  \centering
  \includegraphics[width=0.8\textwidth]{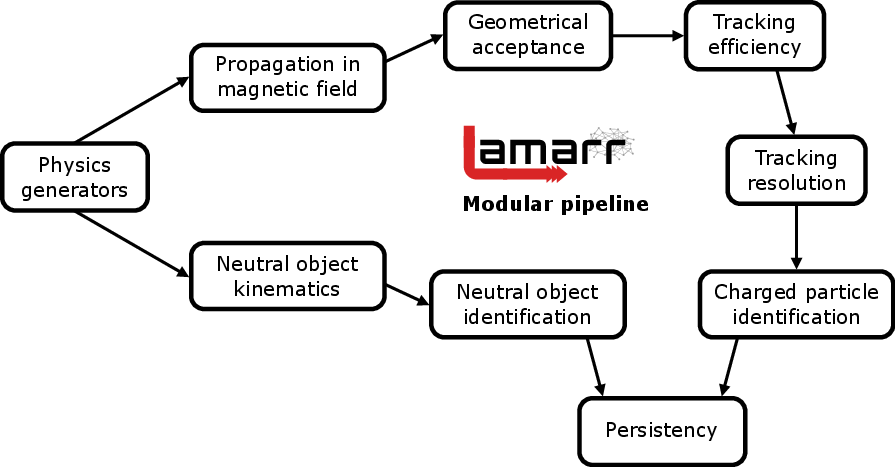}
  \caption{
  Scheme of the \lamarr modular pipeline, illustrating the distinct parameterization paths for charged and neutral particles. 
}
  \label{fig:Lamarr_pipeline}
\end{figure}

Flash simulations provide the fastest simulation options by directly parameterizing the high-level response of the detector.
\textsc{Lamarr}~\cite{FastSimulations:Lamarr,FastSimulations:Lamarr2,FastSimulations:Lamarr3} is the in-house, flash simulation framework for \lhcb consisting of a pipeline of modular ML-based parameterizations, many of which are based on machine learning algorithms.
It starts by processing the particle information from physics generators in \gauss and outputs the high-level response of LHCb sub-detectors.
The pipeline is divided into two chains and is illustrated in Figure~\ref{fig:Lamarr_pipeline}.
The first one targets charged particles and includes tracking acceptance, efficiency and resolution, as well as particle identification.
The second chain is designed for neutral particles, where calorimeters play a key role.

In \lhcb, the momentum of charged particles is measured by exploiting their deflection in the dipole magnet field.
\textsc{Lamarr} parameterizes particle trajectories using the single transverse momentum kick approximation, modeling them as two rectilinear segments with a deflection point inversely proportional to the transverse momentum.
Feed-forward dense neural networks are trained to predict the geometrical acceptance of tracks and tracking efficiency based on kinematics and particle species. The parametrization of tracking efficiency is modeled after a multi-category classification task, extending its validity to particles being detected in a subset of the LHCb tracking detectors, only.
Generative Adversarial Networks (GAN) are employed to simulate resolution effects, such as multiple scattering, and to model the Kalman filter's correlation matrix used in track reconstruction.
These parameterizations enable \textsc{Lamarr} to provide high-level tracking responses, which can be further processed using \lhcb analysis software to reconstruct decay candidates.

Particle identification is crucial for many \lhcb physics analyses to discriminate between different particle species such as muons, pions, kaons, and protons.
\textsc{Lamarr} provides GAN-based parameterizations, conditioned on particle species, kinematics, and detector occupancy.
The GlobalPID variables, combining responses from RICH, MUON, and a loose muon-identification binary criterion implemented at hardware-trigger level, are also parameterized using conditioned GANs with Wasserstein distance loss and a Lipschitz-constrained discriminator.

\textsc{Lamarr} currently employs a simplified calorimeter parameterization for detector studies.
However, the assumption underlying the parametrizations for charged particles, that a one-to-one mapping between the generated particle and reconstructed object exists, does not hold for calorimeters: photons from $\pi^0$ decays can merge into a single cluster, while a single electron may emit multiple Bremsstrahlung photons resulting in as many clusters.
To address this, Graph Attention Networks (GATs) and Transformer architectures are being explored to model the calorimeter response, leveraging attention mechanisms to capture complex correlations.

In order to integrate \textsc{Lamarr} within the \lhcb software stack, parameterizations need to be queried from a C++ application running in the \gaudi framework.
To avoid overheads from multi-threading schedulers, models trained with \textsc{scikit-learn} and \textsc{Keras} are converted to C code using \textsc{scikinC} and distributed via \textsc{cvmfs}.
The \textsc{SQLamarr} package provides low-level components with minimal dependencies that are being integrated within \textsc{Gaussino} aiming at an experiment-independent ultra-fast simulation framework. A proof-of-concept for a stand-alone deployment in the Python ecosystem, named \textsc{PyLamarr}, is also available.

The physics validation of \textsc{Lamarr} is performed by comparing the distributions of ML models trained on detailed simulations with those of standard simulation strategies.
Validation studies using $\Lambda_b^0 \to \Lambda_c^+ \mu^- \bar{\nu}_\mu$ decays, with $\Lambda_c^+ \to pK^-\pi^+$, and $B^+ \to \chi_{c1} K^+$ with $\chi_{c1} \to J\!/\!\psi \gamma$, demonstrate that the decay dynamics and resolution effects are well reproduced, while misreconstruction effects in the neutral sector escape current parametrizations.
\textsc{Lamarr} achieves a two-order-of-magnitude CPU reduction for the simulation phase compared to \geant-based production, with \pythia becoming the major resource consumer.
As parametrizations account for multiplicity effects, a further speed-up can be achieved by simulating signal-only events, with negligible effect on physics performance.

\section{Summary}

Fast simulations aim to replace computationally intensive parts of the simulation with parameterizations that maintain a realistic representation of the detector response.
The \textsc{CaloML} framework, based on generative models, achieves up to two orders of magnitude faster simulation for electrons and photons compared to traditional methods.
Using models such as VAE and their modifications, \textsc{CaloML} provides accurate energy deposit predictions.
Physics validation demonstrates very good agreement between fast and detailed simulations on reconstructed observables, ensuring high fidelity for reconstructed events.
Future advancements, such as the adoption of more sophisticated models such as \textsc{CaloDiT}~\cite{CaloDiT}, could further enhance the realism and precision of fast simulation samples.

Flash simulations directly parameterize the high-level detector response, offering ultra-fast solutions for simulation.
\textsc{Lamarr} employs modular ML-based parameterizations for tracking, particle identification, and calorimeter responses, achieving significant CPU time reductions.
Validation studies confirm the accuracy of its parameterizations.
Ongoing efforts focus on addressing challenges in neutral particle simulation, such as particle-to-particle correlations, using advanced architectures like GNNs and Transformers.
The integration of \textsc{Lamarr} with the LHCb simulation framework and its potential availability to the broader HEP community are key areas of future development.

\bibliographystyle{bib/JHEP}
\bibliography{bib/LHCb-PAPER,bib/LHCb-DP,bib/LHCb-TDR,bib/standard,bib/custom}

\end{document}